\documentclass[twocolumn, twocolappendix]{aastex7}

\usepackage{amsmath}
\usepackage{microtype}
\usepackage{xspace}
\usepackage{makecell}

\newcommand{\comment}[1]{}

\begin{document}

\title{Quantifying the Impact of LSST $u$-band Survey Strategy\\on Photometric Redshift Estimation and the Detection of Lyman-break Galaxies}

\correspondingauthor{John Franklin Crenshaw}
\email{jfc20@uw.edu}

\author[0000-0002-2495-3514]{John Franklin Crenshaw}
\affiliation{Department of Physics, University of Washington, Box 351560, Seattle, WA 98195, USA}
\affiliation{DIRAC Institute, Department of Astronomy, University of Washington, Box 351580, Seattle, WA 98195, USA}
\email{jfc20@uw.edu}

\author[0000-0002-3962-9274]{Boris Leistedt}
\affiliation{Department of Physics, Imperial College London, Blackett Laboratory, Prince Consort
Road, London SW7 2AZ, UK}
\email{b.leistedt@imperial.ac.uk}

\author[0000-0002-9154-3136]{Melissa Lynn Graham}
\affiliation{DIRAC Institute, Department of Astronomy, University of Washington, Box 351580, Seattle, WA 98195, USA}
\email{mlg3k@uw.edu}

\author[0000-0002-1818-929X]{Constantin Payerne}
\affiliation{Université Paris-Saclay, CEA, IRFU, 91191, Gif-sur-Yvette, France}
\email{constantin.payerne@gmail.com}

\author[0000-0001-5576-8189]{Andrew J. Connolly}
\affiliation{DIRAC Institute, Department of Astronomy, University of Washington, Box 351580, Seattle, WA 98195, USA}
\affiliation{eScience Institute, University of Washington, Box 351570, Seattle, WA 98195, USA}
\email{ajc@astro.washington.edu}

\author[0000-0003-1530-8713]{Eric Gawiser}
\affiliation{Department of Physics and Astronomy, Rutgers, the State University of New Jersey, Piscataway, NJ 08854, USA}
\email{gawiser@physics.rutgers.edu}

\author[0000-0002-5652-8870]{Tanveer Karim}
\affiliation{David A. Dunlap Department of Astronomy \& Astrophysics, University of Toronto, 50 St George Street, Toronto, ON, M5S 3H4, Canada}
\affiliation{Dunlap Institute for Astronomy \& Astrophysics, University of Toronto, 50 St George Street, Toronto, ON, M5S 3H4, Canada}
\email{tanveer.karim@utoronto.ca}

\author[0000-0002-8676-1622]{Alex I. Malz}
\affiliation{McWilliams Center for Cosmology and Astrophysics, Department of Physics, Carnegie Mellon University, Pittsburgh, PA, USA}
\email{aimalz@nyu.edu}

\author[0000-0001-8684-2222]{Jeffrey A. Newman}
\affiliation{Department of Physics and Astronomy and PITT PACC, University of Pittsburgh, Pittsburgh, PA 15260, USA }
\email{janewman@pitt.edu}

\author[0000-0002-3645-9652]{Marina Ricci}
\affiliation{Université Paris Cité, CNRS, AstroParticule et Cosmologie, F-75013 Paris, France}
\email{ricci@apc.in2p3.fr}

\collaboration{100}{The LSST Dark Energy Science Collaboration}

\begin{abstract}
    The Vera C. Rubin Observatory will conduct the Legacy Survey of Space and Time (LSST), promising to discover billions of galaxies out to redshift 7, using six photometric bands ($ugrizy$) spanning the near-ultraviolet to the near-infrared.
    The exact number of and quality of information about these galaxies will depend on survey depth in these six bands, which in turn depends on the LSST survey strategy: i.e., how often and how long to expose in each band.
    $u$-band depth is especially important for photometric redshift (photo-$z$) estimation and for detection of high-redshift Lyman-break galaxies (LBGs).
    In this paper we use a simulated galaxy catalog and an analytic model for the LBG population to study how recent updates and proposed changes to Rubin's $u$-band throughput and LSST survey strategy impact photo-$z$ accuracy and LBG detection.
    We find that proposed variations in $u$-band strategy have a small impact on photo-$z$ accuracy for $z < 1.5$ galaxies, but the outlier fraction, scatter, and bias for higher redshift galaxies vary by up to 50\%, depending on the survey strategy considered.
    The number of $u$-band dropout LBGs at $z \sim 3$ is also highly sensitive to the $u$-band depth, varying by up to 500\%, while the number of $griz$-band dropouts is only modestly affected.
    Under the new $u$-band strategy recommended by the Rubin Survey Cadence Optimization Committee, we predict $u$-band dropout number densities of $110$\,deg$^{-2}$ (3200\,deg$^{-2}$) in year 1 (10) of LSST.
    We discuss the implications of these results for LSST cosmology.
\end{abstract}

%\tableofcontents

%\clearpage

\section{Introduction}
\label{sec:intro}

\begin{figure*}
    \centering
    \includegraphics[width=0.85\linewidth]{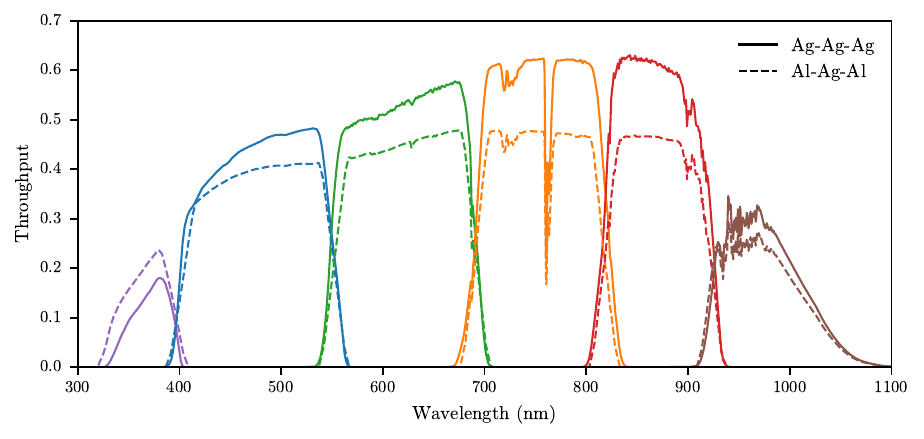}
    \caption{
        Comparison of Rubin Observatory $ugrizy$ throughput curves, assuming original Al-Ag-Al and new Ag-Ag-Ag mirror coatings.
        The transition to all-silver coatings decreased throughput in the $u$ band, but increased throughput in all of the $grizy$ bands.
        These curves include contributions from the atmosphere (assuming airmass 1.2), mirror reflectivities, lens and filter throughputs, and detector sensitivity.
    }
    \label{fig:throughputs}
\end{figure*}

The Vera C. Rubin Observatory's Legacy Survey of Space and Time (LSST) will survey approximately 18,000\,deg$^{-2}$ of the southern sky to unprecedented depth across such a wide area \citep{ivezic2019}.
This is made possible by an 8.4\,m primary mirror and 3.2 Gigapixel camera, yielding an etendue significantly larger than any other existing telescope \citep{lsstScienceBook}.
Six photometric bands spanning the near-ultraviolet to near-infrared will provide information about the spectral energy distributions of objects detected by LSST, enabling, among other things, photometric redshift (photo-$z$) estimation for billions of galaxies.
The quality of this information, however, depends on the depth in each of these bands, which in turn depends on the survey strategy.
Of particular importance for static science is the number and length of visits to each field in the survey footprint.

Over many years, the Rubin Survey Cadence Optimization Committee (SCOC), in collaboration with the Rubin science community, has refined LSST survey strategy to maximize Rubin's science potential, while balancing the needs of a diverse set of science cases (e.g., \citealt{marshall2017, jones2021, scoc_phase1, scoc_phase2, scoc_phase3, bianco2022, lochner2022}).
This is an ongoing process that will continue throughout the 10 year duration of LSST.
Recently, the decision was made to change the coating of Rubin's primary (M1),  secondary (M2), and tertiary mirrors (M3)\footnote{Rubin's primary and tertiary mirrors are a singular structure, usually referred to as the M1M3.} from aluminum-silver-aluminum (Al-Ag-Al) to silver-silver-silver (Ag-Ag-Ag)\footnote{Note this decision was made before the mirrors were ever coated, so the M1M3 mirror was never actually coated with aluminum.}.
This change reduces the throughput in the ultraviolet (i.e., in Rubin's $u$ band), while increasing throughput at longer wavelengths (Rubin's $grizy$ bands; see Fig.~\ref{fig:throughputs}).
Because substantially more survey time is spent observing in $grizy$ bands than in the $u$ band alone, this results in greater survey efficiency, enabling deeper imaging by the end of LSST.
Under the nominal survey strategy at the time of this change, however, the change in mirror coating reduced the 10-year $u$-band depth by 0.21 magnitudes.

The $u$ band, spanning approximately 3300 - 4000\,\AA, is important for a variety of science cases.
At low redshifts, the redshifting of the $\sim 4000$\,\AA~ Balmer break through the $u$ band enables accurate photo-$z$ estimation for galaxies at redshifts $z < 0.5$ \citep{kalmbach2020}.
Without the $u$ band, photo-$z$'s for these galaxies are highly uncertain, resulting in catastrophic outliers that significantly degrade photo-$z$ estimation up to redshifts $z < 0.6$ \citep{lsstScienceBook}.
This has implications for nearly every area of LSST cosmology, including lens and source selection for galaxy clustering and lensing analyses \citep{schmidt2020, zuntz2021, zhang2023, leonard2024}, galaxy cluster detection \citep{adam2019}, supernova cosmology \citep{chen2022, chen2024, mitra2023}.
Achieving precision cosmology therefore places stringent constraints on photo-$z$ performance \citep{descSRD, newman2015, newman2022}.

At higher redshifts, Lyman-series absorption lines shift into the $u$ band, beginning with the Lyman-alpha transition at rest-frame 1216\,\AA, which redshifts into the $u$ band at $z \sim 1.7$, and culminating with the Lyman-limit at rest-frame $912$\,\AA, which redshifts into the $u$ band at $z \sim 2.6$.
These features, caused by absorption from neutral hydrogen in the intergalactic medium (IGM) and within the galaxies themselves, enable identification and photo-$z$ estimation for high-redshift star-forming galaxies known as Lyman-break galaxies (LBGs; \citealt{steidel1996}).

For thirty years, LBGs have been important for studies of galaxy formation and evolution \citep{giavalisco2002, dunlop2013}, including the ultra-high-redshift LBGs discovered in recent years by JWST, illuminating galaxy formation at cosmic dawn (e.g., \citealt{finkelstein2022, mason2022,harikane2023}).
LBGs, however, are also increasingly being recognized as cosmologically important tracers of the matter density field at high redshifts.
With the width and depth of LSST, using LBGs as cosmological tracers has the potential to provide unprecedented constraints on the growth of structure and evolution of dark energy at redshifts $2 < z < 6$ \citep{wilson2019,miyatake2022}; competitive constraints on the amplitude of local-type primordial non-Gaussianity \citep{schmittfull2018, chaussidon2024}; constraints on the sum of neutrino masses, independent of the optical depth of reionization \citep{yu2018}; constraints on the masses of $z > 1$ clusters \citep{tudorica2017}; and constraints on the low-redshift matter density field via inverse galaxy-galaxy lensing (IGGL; \citealt{cross2024}).

The $u$-band dropout technique, which selects galaxies with a strong $u$-band flux deficit compared to the $g$- or $r$-band flux, is especially effective at selecting LBGs at redshifts $2 < z < 4$.
The success of this technique is strongly dependent on $u$-band depth.
By selecting $u$-band dropouts in images from the CFHT Large Area $u$-band Survey (CLAUDS; \citealt{sawicki2019}) and deep $grz$ imaging from Hyper Suprime Cam Subaru Strategic Program (HSC SSP; \citealt{aihara2019}), \citet{ruhlmann-kleider2024} demonstrated it is possible to achieve a spectroscopically-confirmed $2.3 < z < 3.5$ LBG number density of 620 deg$^{-2}$, from an initial photometric sample of $1100$\,deg$^{-2}$ with $r < 24.2$.
The 10-year LSST Wide Fast Deep (WFD) survey, however, will be shallower than the CLAUDS and HSC SSP deep fields.
More recently, \citet{payerne2024} employed a refined LBG selection method on shallower imaging, simulating the ongoing Ultraviolet Near Infrared Optical Northern Survey (UNIONS, \citealt{gwyn2025}) which will have a depth similar to LSST year 2.
This work achieved a confirmed LBG number density of $493$\,deg$^{-2}$ within $2 < z < 3.5$ from a photometrically-selected sample of $1100$\,deg$^{-2}$ with $r < 24.3$, providing valuable insights into LSST's near-future potential for probing dark energy, growth of structure, and primordial non-Gaussianity.

It is therefore essential to understand how LSST survey strategy impacts $u$-band depth and the corresponding implications for photo-$z$ estimation and high-redshift cosmology.
Due to the increased survey efficiency provided by the Ag-Ag-Ag mirror coatings, it is possible to allocate observing time in each band such that the coadded depths in all bands are deeper than the nominal depths assuming the original Al-Ag-Al mirror coatings.
The question, therefore, is how to best balance the depths in each of the six bands to optimize photo-$z$ estimation and LBG detection.

In this paper, we use simulations and simple analytic models to address these questions.
Section~\ref{sec:opsim} details the LSST survey strategy simulations used in this paper.
Section~\ref{sec:photo-$z$} uses these simulations to study how $u$-band observing strategy impacts photo-$z$ estimation, while Section~\ref{sec:lbgs} studies the impact on detection of LBGs.
We discuss the implications of our results for LSST cosmology and conclude in Section~\ref{sec:conclusion}.

We assume \citet{planck2018} cosmology throughout.
We use AB magnitudes, and refer to apparent magnitudes, N-sigma extinction-corrected point-source depths, and cuts in LSST bands as $u$, $u_N$, $u_\text{cut}$, etc.
We use lowercase $m$ to refer to apparent magnitudes in arbitrary bands (i.e. any of $ugrizy$), and uppercase $M$ to refer to absolute magnitudes at rest-frame 1500\,\AA.
For each LBG dropout sample, we use ``dropout band'' to refer to the Rubin band in which the redshifted Lyman-break falls, and ``detection band'' to refer to the band closest to rest-frame 1500\,\AA.
Specifically, for $ugriz$ dropout samples, the dropout bands are $ugriz$ and the detection bands are $rizzy$.
This notation is summarized in Table~\ref{tab:notation}.

Finally, we note this paper contributes to an ongoing literature studying the impact of survey strategy on photo-$z$ estimation, including \citet{graham2018,lochner2018,graham2020,malz2021,lochner2022,scott2024,hang2024}.

\begin{deluxetable}{ll}[t]
    \tablewidth{\linewidth}
    \tablecaption{Summary of notation used in paper}
    \tablehead{\colhead{Notation} & \colhead{Description}} 
    \startdata
        $u,g,r,i,z,y$ & Apparent magnitudes in LSST bandpasses \\
        $m$ & Apparent magnitude in an arbitrary band \\
        $m_N$ & N$\sigma$ extinction-corrected point-source depths \\
        $m_\text{cut}$ & Cut applied to apparent magnitude \\
        $M$ & Absolute magnitude at rest-frame 1500\,\AA
        \rule[-1.8ex]{0pt}{0pt} \\
        \hline
        \rule{0pt}{4.5ex}\!\!
        \makecell[l]{``Dropout band''\\\,} & \makecell[l]{Bandpass in which the Lyman-break falls \\ (i.e., containing $(1+z)\,912$\,\AA)} \\
        \makecell[l]{``Detection band''\\\,} & \makecell[l]{Bandpass closest to rest-frame 1500\,\AA \\ (i.e., closest to $(1+z)\,1500$\,\AA)}
        \rule[-3ex]{0pt}{0pt} \\
    \enddata
    \tablecomments{
        All magnitudes are in the AB system;
        for $ugriz$-dropouts, the detection bands are $rizzy$, respectively.
    }
    \label{tab:notation}
    \vspace{-0.69cm}
\end{deluxetable}

%\section{Simulations and IGM\,$+$\,LBG Models}
%\label{sec:simulations}

\begin{deluxetable*}{lrrrrrrr}
    \tablewidth{\linewidth}
    \tablecaption{$u$-band strategy variations to baseline~v3.4}
    \tablehead{
        \colhead{Strategy} & 
        \colhead{\hspace{0.1cm}\thead{Relative\\$u$ time}} & 
        \colhead{\hspace{0.5cm}$\Delta u$} & 
        \colhead{\hspace{0.5cm}$\Delta g$} & 
        \colhead{\hspace{0.5cm}$\Delta r$} & 
        \colhead{\hspace{0.5cm}$\Delta i$} & 
        \colhead{\hspace{0.5cm}$\Delta z$} & 
        \colhead{\hspace{0.5cm}$\Delta y$}
    } 
    \startdata
        1.0x, 30s $u$ & 1.00 &  0.00  (0.00) &  0.00 ( 0.00) &  0.00 ( 0.00) &  0.00 ( 0.00) &  0.00 ( 0.00) &  0.00 ( 0.00) \\
        1.0x, 38s $u$ & 1.27 &  0.20  (0.18) &  0.00 (-0.01) & -0.04 (-0.01) & -0.02 (-0.01) &  0.02 (-0.01) & -0.01 (-0.00) \\
        1.0x, 45s $u$ & 1.50 &  0.32  (0.31) & -0.03 (-0.01) & -0.05 (-0.02) & -0.03 (-0.02) &  0.00 (-0.02) & -0.02 (-0.01) \\
        1.0x, 60s $u$ & 2.00 &  0.56  (0.52) & -0.08 (-0.04) & -0.07 (-0.04) & -0.05 (-0.03) & -0.02 (-0.04) & -0.04 (-0.02) \\
        1.1x, 30s $u$ & 1.10 &  0.02  (0.04) & -0.01 (-0.01) & -0.02 (-0.00) & -0.01 (-0.01) &  0.02 ( 0.00) & -0.00 (-0.00) \\
        1.1x, 38s $u$ & 1.39 &  0.20  (0.24) & -0.01 (-0.02) & -0.03 (-0.01) & -0.02 (-0.01) &  0.01 (-0.01) & -0.01 (-0.00) \\
        1.1x, 45s $u$ & 1.65 &  0.31  (0.36) & -0.04 (-0.02) & -0.04 (-0.02) & -0.02 (-0.02) & -0.00 (-0.02) & -0.01 (-0.01) \\
        1.1x, 60s $u$ & 2.20 &  0.56  (0.55) & -0.07 (-0.04) & -0.09 (-0.04) & -0.05 (-0.04) & -0.02 (-0.04) & -0.04 (-0.02) \\
        1.2x, 30s $u$ & 1.20 &  0.04  (0.09) & -0.01 (-0.01) & -0.02 (-0.01) & -0.01 (-0.01) &  0.03 (-0.01) & -0.01 (-0.01) \\
        1.2x, 38s $u$ & 1.52 &  0.25  (0.29) &  0.00 (-0.03) & -0.03 (-0.02) & -0.03 (-0.02) & -0.01 (-0.02) & -0.01 (-0.01) \\
        1.2x, 45s $u$ & 1.80 &  0.37  (0.41) & -0.03 (-0.03) & -0.06 (-0.02) & -0.04 (-0.03) &  0.02 (-0.03) & -0.03 (-0.02) \\
        1.2x, 60s $u$ & 2.40 &  0.56  (0.61) & -0.09 (-0.05) & -0.07 (-0.04) & -0.06 (-0.05) & -0.03 (-0.05) & -0.05 (-0.04) \\
        1.5x, 30s $u$ & 1.50 &  0.10  (0.22) & -0.00 (-0.01) & -0.03 (-0.02) & -0.03 (-0.04) &  0.01 (-0.01) & -0.00 (-0.01) \\
        1.5x, 38s $u$ & 1.90 &  0.28  (0.40) & -0.02 (-0.04) & -0.04 (-0.04) & -0.02 (-0.04) &  0.00 (-0.03) & -0.01 (-0.02) \\
        1.5x, 45s $u$ & 2.25 &  0.41  (0.53) & -0.06 (-0.05) & -0.07 (-0.04) & -0.05 (-0.05) & -0.02 (-0.04) & -0.03 (-0.03) \\
    \enddata
    \tablecomments{
        The second columns quantifies the time spent observing in the $u$ band, relative to the baseline~v3.4 observing strategy.
        The right six columns display changes in year 1 (10) $ugrizy$ median 5$\sigma$ depths for the WFD survey.
        For comparison, the median 5$\sigma$ depths for the [1.0x, 30s $u$] strategy are 24.04 (25.15), 25.48 (26.65), 25.69 (26.84), 25.29 (26.40), 24.53 (25.73), 23.68 (24.78) for $ugrizy$ in year 1 (10), respectively.
        Note that the [1.0x, 30s $u$] strategy is the same as baseline~v3.4.
    }
    \label{tab:u-band-sims}
    \vspace{-0.69cm}
\end{deluxetable*}

\comment{
Different survey strategies are simulated using the Rubin Observatory Operations Simulator, described in Section~\ref{sec:opsim}.
The impact on photo-$z$ estimation is studied using a simulated galaxy catalog, which is described in Section~\ref{sec:catalog}, including modeling the effects of IGM extinction in the $u$ and $g$ bands.
High-redshift LBGs are studied using an analytic model, described in Section~\ref{sec:lbg-pop-model}, which enables consistent comparison between observing strategies and obviates the need for a simulated high-redshift catalog.
}

\section{Rubin Operations Simulator}
\label{sec:opsim}

The Rubin Observatory Operations Simulator\footnote{\url{https://rubin-sim.lsst.io/}} (OpSim) generates mock observations over the 10-year duration of LSST.
These simulations include telescope movement, dithering, variable weather and seeing conditions drawn from a Cerro-Tololo Inter-American Observatory (CTIO) historical log, and simulated downtime due to weather and maintenance.
OpSim outputs are processed by the Metrics Analysis Framework (MAF; \citealt{maf}) which computes spatially-varying summary statistics (e.g. median seeing) and derived metrics (e.g. coadded $5\sigma$ depths) that are used to assess survey strategy with regards to survey efficiency and various science drivers.

We primarily focus on the LSST Wide Fast Deep (WFD) survey, which excludes areas with high galactic extinction, $E(B-V)>0.2$.
We use maps of extragalactic extinction-corrected coadded $5\sigma$ point-source depths: $\{u_5, g_5, r_5, i_5, z_5, y_5\}$.
These maps are in  HEALPix \citep{gorski2005} format with $N_{\rm side}=128$, corresponding to a pixel size of 755\,arcmin$^2$.

\begin{figure*}
    \centering
    \includegraphics[scale=1.2]{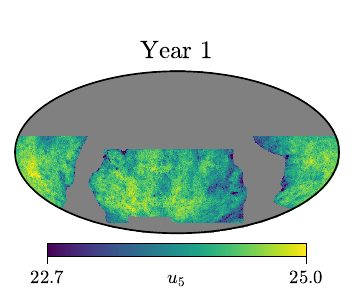}
    \hspace{0.5cm}
    \includegraphics[scale=1.2]{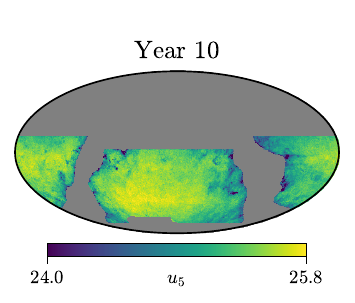}
    \caption{
        Maps of extinction-corrected coadded $5\sigma$ point-source depth in the LSST $u$ band, assuming survey strategy baseline~v4.0. 
        The left panel displays the depth map for LSST year 1, while the right displays year 10.
        By year 10 the $u$-band depth is much deeper (notice the change in color bar limits) as well as significantly more uniform.
    }
    \label{fig:u-depth-maps}
\end{figure*}

To quantify the impact of $u$-band strategy, we compare a series of recent ``baseline'' simulations,
\begin{itemize}
    \item baseline~v3.4: fiducial simulation of LSST including throughputs for the Ag-Ag-Ag mirror coatings, using band allocations identical to the most recent Al-Ag-Al fiducial simulation \citep{opsim3p4}.
    \item baseline~v3.5: an update to baseline~v3.4 that includes 10\% more visits with 38 second exposures in the $u$ band (compared to previous 30 second exposures), as well as the uniform rolling strategy (see \citet{scoc_phase3}) \citep{opsim3p5}.
    \item baseline~v3.6: an update to baseline~v3.5 that includes a more realistic estimate of observatory downtime in year 1 (8 weeks), reducing the number of visits by $\sim 5\%$, as well as an improved model of mechanical inefficiencies in telescope slewing \citep{opsim3p6}.
    \item baseline~v4.0: an update to baseline~v3.6 that includes minor bugfixes to the year 1 downtime that result in slightly more observation time in year 1 \citep{opsim4p0}.
\end{itemize}
For illustration, $u_5$ maps for LSST years 1 and 10 are shown in Fig.~\ref{fig:u-depth-maps}, from which it can be seen, as expected, the $u$-band depth is much deeper and significantly more uniform in year 10 than in year 1.

In addition to these baseline strategies, we analyze a suite of variations to baseline~v3.4 that vary only the $u$-band observing allocation, while simultaneously adjusting the $grizy$ allocations to maintain the 10-year duration of LSST (i.e. increasing $u$-band observing time necessitates decreasing $grizy$ observing time, which are each decreased by the same proportion).
These strategies include increasing the number of $u$-band visits by 10\%, 20\%, and 50\%, while setting the per-visit exposure time to 30, 38, 45, and 60 seconds.
Throughout, we refer to these strategies with names such as [1.1x, 38s $u$], which refers to the strategy that increases $u$-band visits by 10\% and the per-visit exposure time to 38 seconds.
Table~\ref{tab:u-band-sims} lists each of these simulated variations, their relative change in overall $u$-band exposure time, and their impact on median depth in each band.
Note that for the same amount of $u$-band observing time, strategies with longer per-visit exposures yield greater depth in the $u$ band compared to more visits with shorter exposures.
This is because noise in the $u$ band is dominated by read noise and not sky background due to the fainter sky background in the ultraviolet.
Also note that the [1.0x, 30s $u$] strategy is the same as baseline~v3.4.
Baselines v3.5 onward have adopted the [1.1x, 38s $u$] strategy, partially as a result of the findings presented here, but these simulations are not identical to the baseline v3.4 [1.1x, 38s $u$] simulation due to the other changes described above.

Finally, in addition to the WFD forecasts we make for every survey strategy, for the baseline v4.0 simulation we also estimate LBG number densities in the LSST Deep Drilling Fields (DDFs): COSMOS, the Extended Chandra Deep Field South (ECDFS), ELAIS S1, XMM-LSS, and the Euclid Deep Field South (EDFS).
The DDFs are typically 1.3 magnitudes deeper in each band than the WFD survey, however EDFS is shallower than the others as it receives approximately the same number of visits spread over roughly twice the area. 
COSMOS, furthermore, has an accelerated schedule to build significant depth within the first 3 years to aid cosmology systematics calibration and low-surface-brightness science.

\section{Impact on photo-$z$ Estimation}
\label{sec:photo-$z$}

In this section we study the impact of the $u$-band observing strategy on photo-$z$ estimation for galaxies at redshifts $z \lesssim 3$.
In Section~\ref{sec:catalog} we discuss our simulated galaxy catalog and formalism for modeling IGM extinction,
in Section~\ref{sec:cmnn} we present our photo-$z$ estimator,
and in Section~\ref{sec:photoz-results} we present our results quantifying the impact of $u$-band strategy on photo-$z$ accuracy.

\subsection{Simulating IGM Extinction for $z \lesssim 3$ Galaxies}
\label{sec:catalog}

To study the impact of $u$-band observing strategy on photo-$z$ estimation we use a simulated catalog based on the Millennium simulation \citep{springel2005}, using the GALFORM semianalytic galaxy formation model \citep{gonzalez-perez2014} and the lightcone construction techniques described by \citet{merson2013}.
This catalog was designed to model the optical and near-infrared properties, including emission lines, of $z \leq 3$ galaxies detected by LSST.
We apply a magnitude cut of $m_i<25.5$, which is slightly deeper than the DESC gold sample \citep{descSRD} to avoid edge effects in our analysis.

This catalog contains true redshifts and $ugrizy$ magnitudes for 240,000 galaxies\footnote{These true magnitudes were computed for the original Al-Ag-Al throughputs. The new throughputs have nearly identical shapes, however, with only the normalization being substantially different. These true magnitudes are, therefore, still valid for the new throughputs.}.
The model that generated these true magnitudes includes Lyman-series absorption in the atmospheres of these galaxies, but does not include the effects of extinction in the intergalactic medium (IGM), which also absorbs rest-frame UV flux from these galaxies as photons travel through neutral hydrogen clouds along the line of sight.
This effect, commonly named the Lyman-alpha forest, is redshift and wavelength dependent, and therefore provides valuable information for photo-$z$ estimation in addition to the information provided by Lyman-series absorption intrinsic to galactic atmospheres.
As this information redshifts into the Rubin $u$ band at $z > 1.6$ (and the $g$ band at $z > 2.2$), we wish to model this missing IGM absorption to maximize the utility of the $u$ band for photo-$z$ estimation in our simulations.

We use the following model to add IGM extinction at the catalog level.
Consider a galaxy with observed-frame SED $f_\lambda(\lambda) \equiv f_\lambda(\lambda,z)\propto f_\lambda(\lambda/(1+z),z=0)$, observed in a bandpass with transmission\footnote{This is the dimensionless throughput, giving the probability that a photon with wavelength $\lambda$ will be detected.} $R_m(\lambda)$.
In the absence of IGM extinction (e.g., in our simulated catalog), we observe the magnitude
\begin{align}
    m &= -2.5 \log_{10} f + c_m \nonumber \\
    \text{where} \quad
    f &= \int f_\lambda(\lambda) R_m(\lambda) \lambda \, d\lambda,
\end{align}
and $c_m$ is a band-dependent constant.
However, if we include the observed-frame IGM transmission $T(\lambda)$, we observe the magnitude
\begin{align}
    m_\mathrm{wIGM} &= -2.5 \log_{10} f_\mathrm{wIGM} + c_m \nonumber \\
    \text{where} \quad
    f_\mathrm{wIGM} &= \int T(\lambda) f_\lambda(\lambda) R_m(\lambda) \lambda \, d\lambda.
\end{align}

\begin{figure}
    \centering
    \includegraphics[width=0.95\linewidth]{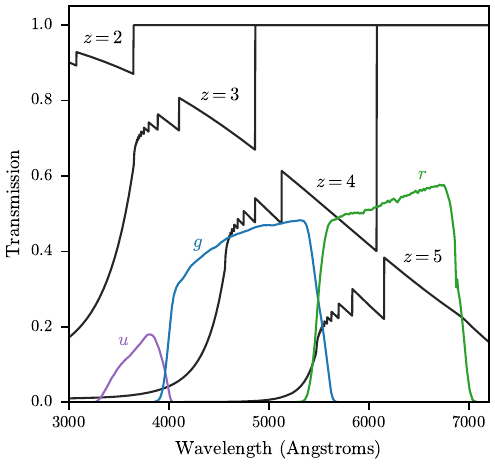}
    \caption{
        Transmission of the \citet{inoue2014} IGM model in black, plotted for several different source redshifts.
        The transmission of the LSST photometric bandpasses (with Ag-Ag-Ag coating) are plotted in color to help visualize how much IGM extinction impacts each band at different redshifts.
    }
    \label{fig:igm}
\end{figure}

We model IGM transmission using the analytic model of \citet{inoue2014}.
This model specifies the optical depth as a function of wavelength and redshift, $\tau(\lambda, z)$, from which IGM transmission can be calculated: $T(\lambda, z) = e^{-\tau(\lambda, z)}$.
Fig.~\ref{fig:igm} plots the three bluest Rubin bandpasses together with the IGM transmission for a few different source redshifts.
The galaxies in our simulated catalog are all at $z \leq 3.5$, for which only the $u$ and $g$ band see any significant IGM extinction.

Absorption in the IGM effectively increments observed magnitudes by the amount
\begin{align}
    \Delta m_\mathrm{wIGM} = m_\mathrm{wIGM} - m = -2.5 \log_{10} \frac{f_\mathrm{wIGM}}{f}.
\end{align}
We wish to compute these IGM corrections for the galaxies in our catalog.
If we assume the UV SED for each galaxy can be approximated by a power law, $f_\lambda(\lambda) \propto \lambda^{\beta_\mathrm{UV}}$ where $\beta_\mathrm{UV}$ is the ``UV slope'', the flux ratio is
\begin{align}
    \frac{f_\mathrm{wIGM}}{f} = \int T(\lambda) \Tilde{R}_m(\lambda) \, d\lambda
\end{align}
where
\begin{align}
    \Tilde{R}_m(\lambda) = \frac{
        \lambda^{\beta_\mathrm{UV} + 1} R_m(\lambda)
    }{
        \int \lambda^{\beta_\mathrm{UV} + 1} R_m(\lambda) \, d\lambda
    }.
\end{align}
The IGM correction $\Delta m_\mathrm{wIGM}$, then, depends only on the UV slope $\beta_\mathrm{UV}$.

For this simple model, we fit $\beta_\mathrm{UV}$ from the $u-g$ color of the galaxies.
Assuming the $u$-band flux is the flux of the galaxy at the effective wavelength of the $u$ band,
\begin{align}
    \lambda_u = \frac{
        \int \lambda \, f_\lambda(\lambda) R_u(\lambda) \lambda \, d\lambda
    }{\int f_\lambda(\lambda) R_u(\lambda) \lambda \, d\lambda}
\end{align}
we have
\begin{align}
    u \sim -2.5 \log_{10} \lambda_{u}^{\beta_\mathrm{UV} + 2} + c_u + c,
\end{align}
where $c$ is a band-independent constant.
Similar equations hold for the $g$ band.
We then estimate the UV slope from the $u-g$ color:
\begin{align}
    \beta_\mathrm{UV} = (u - g) \left( -2.5 \log_{10}\frac{\lambda_u}{\lambda_g} \right)^{-1} - 2.
\end{align}
Note, however, that this definition is circular, as the effective wavelength depends on the UV slope.
We therefore estimate initial effective wavelengths assuming $\beta_\text{UV} = -2$, estimate new values for $\beta_\text{UV}$, and iterate until convergence.

\begin{figure}
    \centering
    \includegraphics[width=\linewidth]{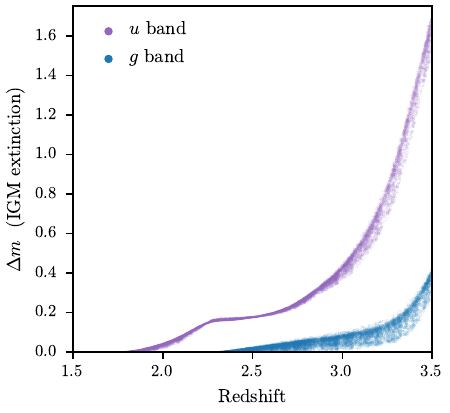}
    \caption{
        IGM magnitude increments in the Rubin $u$ and $g$ bands for $z > 1.5$ galaxies in the simulated catalog.
        The scatter is due to scatter in UV slope, $\beta_\text{UV}$.
    }
    \label{fig:corrections}
\end{figure}

$\Delta m_\mathrm{wIGM}$ for galaxies in the simulated catalog are plotted in Fig.~\ref{fig:corrections}.
The dispersion in $\Delta m_\mathrm{wIGM}$ is due to the UV slopes of the galaxies.
Whether points above (below) the mean are bluer (redder) than average depends on the source redshift.
See Appendix~\ref{sec:beta-dependence} for more details.

For our simulated catalog, final estimates of $\beta_\mathrm{UV}$ range from $-4$ for extreme star formers, to $8$ for very red galaxies.
The bottom of this range matches expectations for extreme star-forming galaxies \citet{bouwens2014, izotov2021}.
We have not found published estimates of UV slopes for red galaxies, as this is a technique typically used for studying star formation.
We note, however, the UV slope provides only a modest modulation to the mean IGM extinction, as seen in Fig.~\ref{fig:corrections}, so we do not expect this simple model to endanger any conclusions of this paper.

\subsection{photo-$z$ Estimation}
\label{sec:cmnn}

We use the Color-Matched Nearest-Neighbors (CMNN) algorithm \citep{graham2018,graham2020} to estimate photo-$z$'s.
CMNN is not chosen because it is the best photo-$z$ estimator, but rather because the accuracy and precision of CMNN estimates are straightforwardly related to the precision of the input photometry.
This attribute makes the CMNN photo-$z$ estimator useful for evaluating the relative change in photo-$z$ performance due to varying photometric quality in different survey strategy simulations.

The CMNN estimator is described fully in \citet{graham2018,graham2020}.
To briefly summarize, CMNN takes only two inputs: the median 5$\sigma$ depths in each of the $ugrizy$ bands for a given survey simulation, and the simulated galaxy catalog of true apparent magnitudes.
Given the depths, CMNN calculates observational magnitude uncertainties for each galaxy (using the error model described in \citealt{ivezic2019,crenshaw2024a}) and uses the errors to simulate observed apparent magnitudes (i.e., adds randomly generated noise to the true flux).
The catalog is then split into a training set of 200,000 galaxies and a test set of 40,000 galaxies, and the training set is used to estimate a photo-$z$ point estimate for each galaxy in the test set.

\subsection{Results: photo-$z$ Estimation}
\label{sec:photoz-results}

\begin{figure*}
    \centering
    \includegraphics[width=0.31\linewidth]{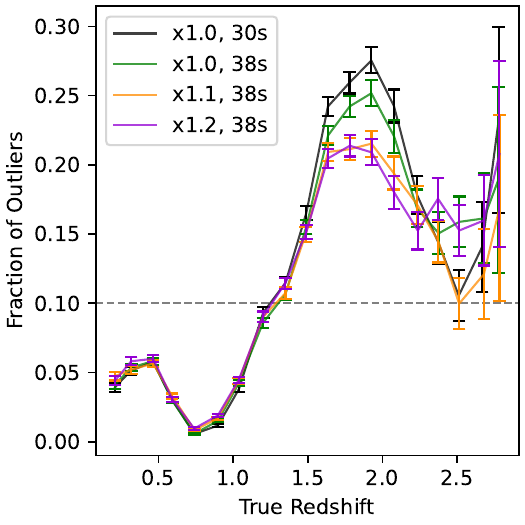}
    \hspace{0.02\linewidth}
    \includegraphics[width=0.31\linewidth]{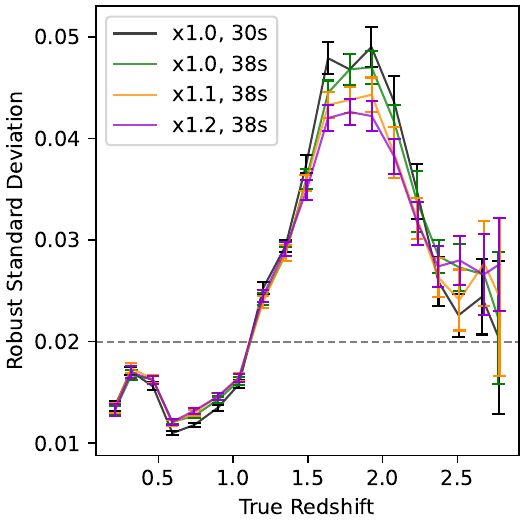}
    \hspace{0.02\linewidth}
    \includegraphics[width=0.31\linewidth]{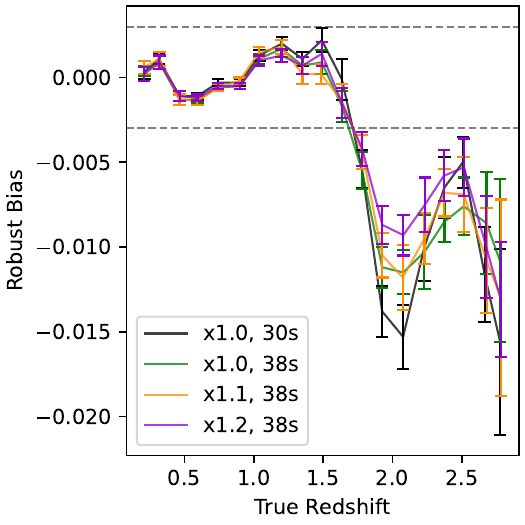}
    \caption{
        photo-$z$ metrics for simulations of several different $u$-band survey strategies.
        Each metric is averaged within 19 overlapping redshift bins of width 0.3, spanning the range $0 \leq z \leq 3$.
        The redshift value for each bin is the mean redshift of galaxies in that bin.
        Uncertainties are estimated by bootstrapping 1000 times.
        For clarity, we plot only a small subset of the $u$-band strategies considered.
        Each panel displays the corresponding requirement for LSST science as a horizontal gray line \citep{lsstSRD}.
        These requirements provide a sense of scale for each metric, but whether or not this photo-$z$ estimator achieves each limit in different redshift ranges is not predictive of the photo-$z$ performance of DESC cosmology, due to the considerations discussed in Section~\ref{sec:cmnn}.
    }
    \label{fig:pz-stats}
\end{figure*}

\begin{figure*}
    \centering
    \includegraphics[width=0.31\linewidth]{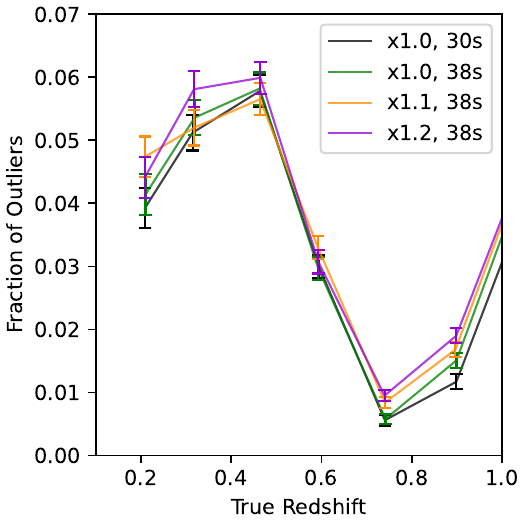}
    \hspace{0.02\linewidth}
    \includegraphics[width=0.31\linewidth]{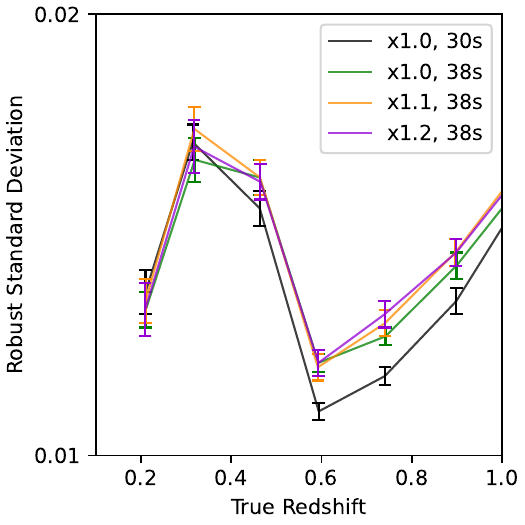}
    \hspace{0.02\linewidth}
    \includegraphics[width=0.31\linewidth]{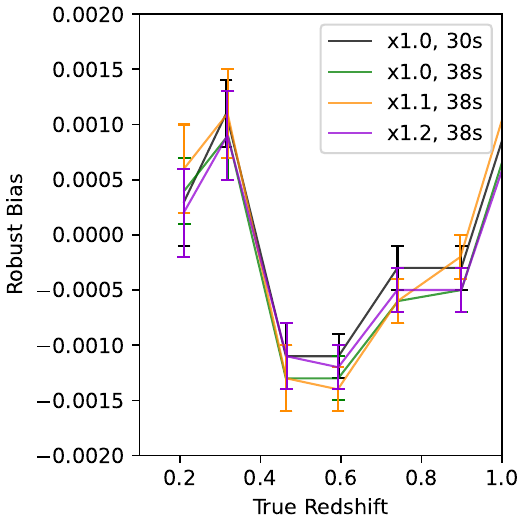}
    \caption{
        Same as Fig.~\ref{fig:pz-stats}, zoomed in on the $z < 1$ region for each metric.
        Recall that the lowest-redshift bin extends to $z=0$ and that the plotted redshift values are the mean redshift for galaxies in each bin.
    }
    \label{fig:pz-stats-zoom}
\end{figure*}

We estimate photo-$z$'s for every galaxy in our simulated catalog using CMNN together with the OpSim maps of $ugrizy$ 5$\sigma$ depths for LSST year 10.
We quantify photo-$z$ accuracy via the quantity $\Delta z = (z_\text{phot} - z_\text{true}) / (1 + z_\text{true})$, the numerator of which quantifies the photo-$z$ error, while the denominator compensates for the naturally larger uncertainty at high redshifts.
We then bin galaxies by true redshift and calculate the following quantities for each bin:
\begin{itemize}
    \item the robust standard deviation, $\sigma_{\Delta z}$, which we define as the width of the interquartile range (IQR) of $\Delta z$, divided by 1.349 to convert to the equivalent of a Gaussian standard deviation;
    \item the photo-$z$ bias, which we define as the mean value of $\Delta z$ for galaxies within the IQR;
    \item the outlier fraction, which we define as the fraction of galaxies for which $|\Delta z| > 3 \sigma_{\Delta z}$.
\end{itemize}
When calculating the first two quantities, we exclude the galaxies that are flagged as outliers, so these two quantities characterize the core of the distribution, while the outlier fraction characterizes the tails.

These quantities, as a function of redshift\footnote{These plots show metrics binned vs true redshifts. Plots of metrics binned vs photo-$z$ are qualitatively the same and do not change our conclusions.}, are plotted in Fig.~\ref{fig:pz-stats}, with the colors corresponding to different $u$-band strategies (recall Section~\ref{sec:opsim} and Table~\ref{tab:u-band-sims}).
Each quantity is averaged within 19 overlapping redshift bins of width 0.3, spanning the range $0 \leq z \leq 3$.
Uncertainty within each bin is estimated by bootstrapping 1000 times.

All three metrics significantly degrade above $z \sim 1.2$, at which point the Balmer break redshifts out of the Rubin bandpasses and there is relatively little information present in galaxy spectra for broadband photo-$z$ estimation, resulting in significantly degraded photo-$z$ accuracy \citep{lsstScienceBook,kalmbach2020}.
Around $z \sim 1.6$, however, Lyman-series transitions (i.e., the Lyman-alpha forest) begin to redshift into the Rubin $u$ band, stealing progressively more of the $u$-band flux.
This provides a distinctive signal for photo-$z$ estimation, resulting in a reversal of the trend, with photo-$z$ accuracy improving until about $z \sim 2.5$.
Our simulations contain very few galaxies beyond this redshift, resulting in increased photo-$z$ errors and uncertainties in the quantities plotted in Fig.~\ref{fig:pz-stats}.
We do not attempt to interpret our results in this noisy regime, and instead defer considerations of higher-redshift galaxies to Section~\ref{sec:lbgs}.

For all three quantities, increasing the total $u$-band exposure time improves photo-$z$ results in the range $1.5 < z < 2.5$:
the outlier fraction improves by up to 30\%;
the standard deviation improves by up to 20\%;
the bias improves by up to 40\%.
These improvements are due to the greater $u$-band depth increasing the SNR of the $u$-band flux decrement that results from the redshifting of Lyman-series absorption into the Rubin $u$ band.

The improvements in photo-$z$ estimation at $1.5 < z < 2.5$, however, come with a loss of performance at $z < 1$ due to the decreased depth in the $grizy$ bands (because more time spent observing in $u$ must result in less time observing in $grizy$, due to the fixed 10-year duration of LSST).
The three panels of Fig.~\ref{fig:pz-stats-zoom} show the same quantities as Fig.~\ref{fig:pz-stats}, zoomed in on the region $z < 1$.
These losses, while smaller in magnitude than the gains at $1.5 < z < 2.5$, are, however, statistically significant, especially for the outlier fraction and scatter, which both degrade by up to 3$\sigma$.
It must also be kept in mind that the $z < 1$ galaxy sample contains far more galaxies, both in our simulations and for the future LSST survey.
Thus these small performance losses at $z < 1$ may be judged to outweigh the gains at $1.5 < z < 2.5$, depending on the science case in consideration.

We note, however, that our simulated photometry, while noisy, is free of biases, aperture corrections, and other systematic errors.
These effects, present in real measured photometry, are likely to erase the small changes in photo-$z$ performance at $z < 1$ that result from the small changes in $grizy$ depth (cf. Table~\ref{tab:u-band-sims}).
We judge, therefore, that the impressive gains at $z > 1.5$ provide strong motivation for increasing $u$-band observing time.
Indeed, baselines v3.5 onward have adopted the [1.1x, 38s u] strategy, partially as a result of the findings presented here.

Finally, in addition to these metrics, it is interesting to specifically consider the number of catastrophic outliers that result from confusion of the Lyman and Balmer absorption features, at roughly 1216\,\AA\ and 3645\,\AA, respectively.
In our $z \lesssim 3$ sample, this results in some $z \sim 0.15$ galaxies being mistaken as $z \sim 2.5$, and vice versa.
A deeper $u$ band enables better discrimination between these two features, reducing the confusion rate.
Compared to the [1.0x, 30s $u$] strategy, the [1.0x, 38s $u$], [1.1x, 38s $u$], [1.2x, 38s $u$] strategies decrease the number of these catastrophic outliers by approximately 6\%, 13\%, and 17\%, respectively.

\section{Impact on LBG Detection}
\label{sec:lbgs}

In this section we study the impact of the $u$-band observing strategy on high-redshift LBG detection.
The mock galaxy catalog used for evaluating photo-$z$ estimation in the previous section does not extend to high enough redshifts to study LBG detection with LSST.
In this section, therefore, we develop an analytic model of the LBG population, which we discuss in Section~\ref{sec:lbg-pop-model}.
In Section~\ref{sec:lbg-forecast} we establish simple criteria for LBG detection,
and in Section~\ref{sec:lbg-results} we present our results quantifying the impact of different $u$-band strategies.

\subsection{LBG Flux and Population Model}
\label{sec:lbg-pop-model}

\begin{figure*}
    \centering
    \includegraphics[width=0.95\linewidth]{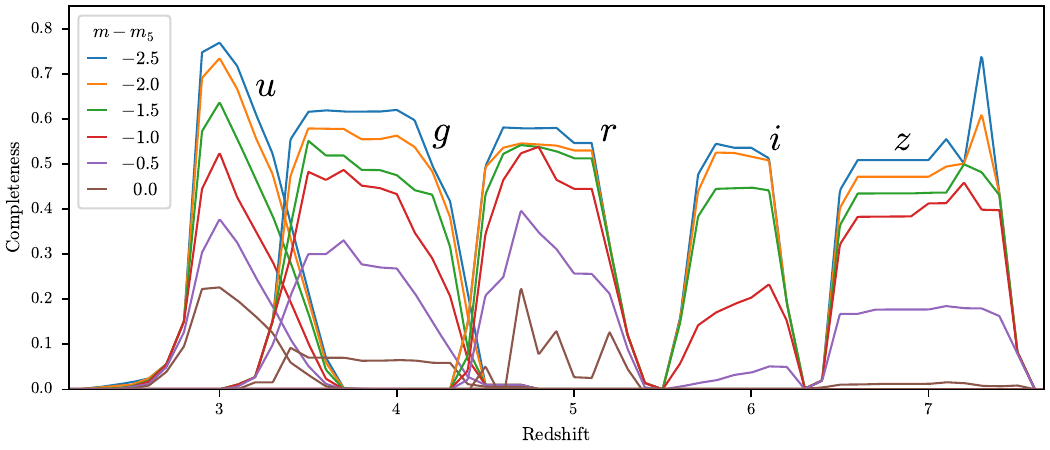}
    \caption{
        LBG completeness curves for $ugriz$ dropouts as a function of redshift, in bins of LBG magnitude relative to the 5$\sigma$ depth in the detection band ($rizzy$, respectively).
        $u$-dropout completeness comes from \cite{malkan2017}; $griz$ completeness comes from \cite{ono2018,harikane2022}.
    }
    \label{fig:completeness}
\end{figure*}

We model the intrinsic rest-frame spectra of LBGs in absolute magnitudes using a power law,
\begin{align}
    f_\lambda^\text{intr}(\lambda, M, z) = A \cdot 10^{-0.4 M} \left( \frac{\lambda}{1500\,\text{\AA}} \right)^{\beta_\text{UV}(M, z)},
    \label{eq:lbg_flux_intr}
\end{align}
where $M$ is the absolute magnitude at 1500\,\AA\footnote{I.e., $M$ is the AB magnitude as calculated with the bandpass $R(\lambda) = \delta(\lambda - 1500\text{\AA})$, where $\delta$ is the Dirac delta function, assuming the galaxy is at 10 pc (for which redshift $z \sim 0$).}, and the normalization $A = 4.83 \times 10^{-8}$\,erg~s$^{-1}$\,cm$^{-2}$\,\AA$^{-1}$.
The UV slope is calculated using the bilinear model
\begin{align}
    \beta_\text{UV}(M, z) = -0.167(M + 19.5) - 0.063 \, z - 1.61,
    \label{eq:beta-model}
\end{align}
which is fit to high-redshift Hubble data from \citet{bouwens2014} (see Appendix~\ref{sec:uv-slope-fit}).
The observed spectrum in apparent magnitudes, accounting for redshift and IGM extinction, is then
\begin{align}
    f_\lambda^\text{obs}(\lambda, M, z) =
    \frac{1}{1 + z}
    \left(
        \frac{10\,\text{pc}}{D_L(z)}
    \right)^{\!2}
    f_\lambda^\text{intr}(\lambda_e, M, z) \,
    T(\lambda, z)
    \label{eq:lbg_flux_obs}
\end{align}
where $D_L(z)$ is the luminosity distance to redshift $z$, and $\lambda_e = \lambda / (1 + z)$ is the emitted wavelength.
Finally, the observed bandpass magnitude is
\begin{align}
    m(M, z) =
    -2.5 \log_{10} \left(
        \frac{
            \int f_\lambda^\text{obs}(\lambda, M, z) R_m(\lambda) \lambda \, d\lambda
        }{
            \int f_\lambda^\text{AB}(\lambda) R_m(\lambda) \lambda \, d\lambda
        }
    \right),
    \label{eq:apparent-mag}
\end{align}
where $f_\lambda^\text{AB}(\lambda) = 0.109 \, (\lambda/\text{\AA})^{-2}$\,erg~s$^{-1}$\,cm$^{-2}$\,\AA$^{-1}$ is the AB reference spectrum (i.e. $f_\nu = 3631$\,Jy).
Note that Equations~\ref{eq:lbg_flux_intr}-\ref{eq:apparent-mag} provide a fully determined analytic model for $m$, the apparent magnitude in bandpass $R_m$, as a function of redshift, $z$, and the absolute magnitude at rest-frame 1500\,\AA, $M$.
Alternatively, by numerically inverting these equations, we have a fully determined model for $M$ as a function of $z$ and $m$.

True LBG number density as a function of redshift and absolute magnitude at 1500\,\AA\, (i.e., the luminosity function) is modeled using a double power law:
\begin{align}
    \phi(M, z) = \phi^* \big[
        &10^{0.4(\alpha + 1)(M - M^*)} \nonumber \\
        &+ 10^{0.4(\beta + 1)(M - M^*)}
    \big]^{-1},
    \label{eq:phi_M_z}
\end{align}
where $\phi^*$ is the characteristic number density, $M^*$ is the characteristic magnitude, $\beta$ is the bright-end slope, and $\alpha$ is the faint-end slope.
For the redshift evolution of each parameter we use the following model:
\begin{align}
    \log \phi^*(z) &= -1.45 - 0.31 (1 + z) \nonumber \\
    M^*(z) &= -21.18 + 0.02 (1 + z) \nonumber \\
    \alpha(z) &= -1.27 - 0.11 (1 + z) \nonumber \\
    \beta(z) &= -4.79 + 0.05 (1 + z).
\end{align}
These coefficients, from Table~3 of \citet{finkelstein2022a}, are fit to a collection of data from CANDELS \citep{finkelstein2015,parsa2016,bouwens2021}, the Hubble Frontier Fields \citep{bouwens2022}, HSC SSP \citep{harikane2022}, CLAUDS \citep{moutard2020}, SHELA \citep{stevans2018}, and UltraVISTA/VIDEO \citep{adams2020}.

The fraction of true LBGs detected and classified as such by Rubin, as a function of redshift, observed magnitude in the detection band, and 5$\sigma$ limiting depth in the detection band, is estimated using the LBG completeness models of \citet{malkan2017,ono2018,harikane2022}.
In particular, the completeness is quantified in terms of $m - m_5$, the magnitude in the detection band, relative to the 5$\sigma$ depth in the same band.
For $ugriz$-dropout samples, we define the detection band as the Rubin bandpass closest to the redshifted $(1+z) 1500$\,\AA: $rizzy$, respectively.
Completeness models, for a discrete set of $m - m_5$ values in the detection band, are shown in Fig.~\ref{fig:completeness}, from which it is clear that the dropout samples are less complete for galaxies with detection band magnitudes closer to the 5$\sigma$ limit.
For more details, see Appendix \ref{sec:completeness_model}.

\begin{figure}
    \centering
    \includegraphics[width=0.98\linewidth]{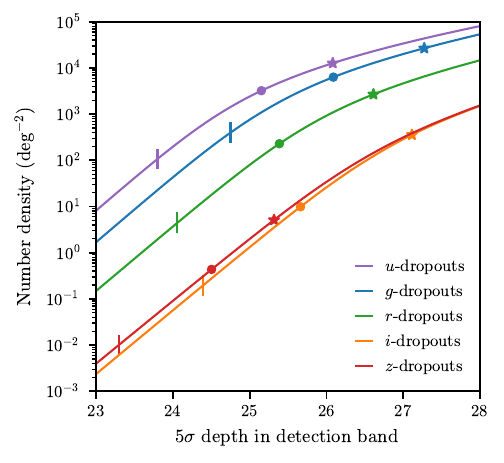}
    \caption{
        Projected number densities for $ugriz$ dropouts as a function of magnitude cut in the detection band, assuming $m_\text{cut} = m_5$.
        For $ugriz$ dropouts the detection band in $rizzy$.
        Expected depths and number densities for the baseline v4.0 survey strategy are indicated by bars, dots, and stars for LSST WFD years 1 and 10, and COSMOS year 10, respectively.
    }
    \label{fig:number_density}
\end{figure}

Finally, we estimate detected LBG number densities by integrating the luminosity function and completeness model:
\begin{align}
    n = \int_0^\infty dz \, \frac{dV}{dz} \int_{-\infty}^{M_\text{cut}} \! dM \, \phi(M, z) \, C(M, z; M_5).
    \label{eq:nz}
\end{align}
The absolute-magnitude cut, $M_\text{cut} = M_\text{cut}(m_\text{cut}, z)$, and the absolute-magnitude $5\sigma$ depth, $M_5 = M_5(m_5, z)$, are calculated from the detection-band apparent magnitude equivalents, $m_\text{cut}$ and $m_5$, by inverting Equation~\ref{eq:apparent-mag}.
Note that $m_5$ is the detection-band apparent magnitude corresponding to a flux with signal-to-noise ratio (SNR) of 5 in the detection band, which is provided by OpSim (see Section~\ref{sec:opsim}).
$m_\text{cut}$, on the other hand, is the cut on detection-band apparent magnitude for LBGs selected for cosmology (see Section~\ref{sec:lbg-forecast}).
Number densities as a function of detection-band $m_5$ (where we have set $m_\text{cut} = m_5$) are shown in Fig.~\ref{fig:number_density}.
Median depths for the baseline v4.0 survey strategy are marked, using bars, dots, and stars for WFD years 1 and 10, and COSMOS year 10, respectively.

We do not model contamination from low-redshift interlopers for the metrics presented in this work.
Low-redshift interlopers, however, are very important for cosmology and astrophysics applications, and are discussed in Section~\ref{sec:conclusion}.

\subsection{Forecasting Dropout Number Densities}
\label{sec:lbg-forecast}

Constraining large-scale structure with LBGs requires assembling a relatively uniform, high-number density sample across the largest possible area of the sky.
These criteria are in tension, as uniformity and number density encourage selection of the deepest areas of the survey, while sky area encourages selecting the widest possible area of the survey (cf. the trade-offs in \citealt{ruhlmann-kleider2024} and \citealt{payerne2024}).
This trade-off between width and depth will need to be optimized for real analyses that seek to maximize cosmological constraining power.
Furthermore, LBG selection criteria, which rely on colors that straddle the observed-frame Lyman-break at $\lambda_\text{obs} \sim 912\,\text{\AA} (1 + z)$, will need to be optimized to balance sample size, completeness, and purity according to the needs of cosmology analyses \citep{wilson2019,petri2025}.
Here, we make a simple set of choices that enables consistent comparison between different observing strategies.

Predicting LBG number densities using Eq.~\ref{eq:nz} requires specifying values for $m_\text{cut}$ (cuts on LBG detection-band apparent magnitudes).
Reasonable choices for $m_\text{cut}$ can be determined if we set the following requirements:
\begin{itemize}
    \item we require detected LBGs have SNR $>5$ in the detection band;
    \item we require a dropout greater than 1 mag, i.e. $m_\text{dropout} - m_\text{detection} \geq 1$;
    \item in the dropout band, magnitudes fainter than the 3$\sigma$ limit are replaced by 3$\sigma$ lower bounds.
\end{itemize}
If we wish to select a single value for $m_\text{cut}$ that is as deep as possible while satisfying these requirements across the entire survey footprint, these requirements translate to
\begin{align}
    m_\text{cut} &= \min \Bigr[
        \min(m_{\text{det}, 5}), ~
        \min(m_{\text{drop}, 3}) - 1
    \Bigr] \\
    &= \min \Bigr[
        \min(m_{\text{det}, 5}), ~
        \min(m_{\text{drop}, 5}) - 0.45
    \Bigr]. \label{eq:mcut}
\end{align}
In the second line we have used
\begin{align}
    m_3 = m_5 + 2.5\log_{10}\frac{5}{3} \approx m_5 + 0.55
\end{align}
and minima are to be taken over pixels in the OpSim maps.

Equation~\ref{eq:mcut} makes clear that $m_\text{cut}$, and therefore the number density of LBGs useful for cosmology, will be limited by the shallowest pixels in the survey footprint.
To mitigate the impact of the shallowest pixels, therefore, we use only the deepest 75\% of the WFD footprint (corresponding to a sky fraction $f_\text{sky} = 0.32$).
This choice balances width and depth in a manner that approximately optimizes the cosmology signal, while yielding a survey footprint that is largely contiguous and mostly overlapping for the different dropout samples.

For $u$- and $g$-dropouts, the dropout band is not sufficiently deep\footnote{Equation~\ref{eq:mcut} makes clear the criterion for being ``sufficiently deep'' is that the 5$\sigma$ depth in the dropout band is more than 0.45 mag deeper than the 5$\sigma$ depth in the detection band. See Table~\ref{tab:u-band-sims}  for the relative depth of each band.} with respect to the detection band ($r$ and $i$, respectively), so that the dropout band is the limiting factor when determining magnitude cuts.
For these samples, therefore, we select the deepest 75\% of pixels in the \textit{dropout band} for each OpSim map, and set the LBG magnitude cuts:
\begin{align}
    u\text{-drop.:} ~~ r_\text{cut} &= \min(u_5) - 0.45 \nonumber \\
    g\text{-drop.:} ~~ i_\text{cut} &= \min(g_5) - 0.45.
    \label{eq:shallow-cuts}
\end{align}

\begin{figure*}[!t]
    \centering
    \includegraphics[width=0.95\linewidth]{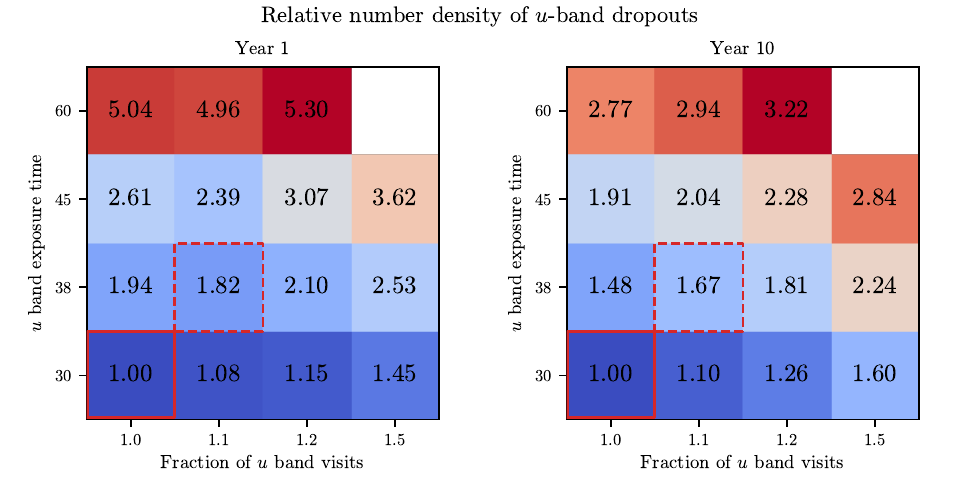}
    \caption{
        Relative number density of $z \sim 3$ $u$-band dropouts detected across the deepest 75\% of the survey footprint for different simulations of $u$-band survey strategy.
        Left panel corresponds to LSST year 1, right panel to year 10.
        The block corresponding to baseline~v3.4, [1.0x, 30s $u$], is bordered by a solid red box, while the SCOC recommendation of [1.1x, 38s $u$] is bordered by a dashed red box.
        The absolute number density for baseline~v3.4 is 69\,deg$^{-2}$ in year 1 and 2113\,deg$^{-2}$ in year 10.
    }
    \label{fig:lbg-grid}
\end{figure*}

For $riz$-dropouts, the dropout band is deeper than the detection band ($zzy$, respectively).
For these samples, therefore, we select the deepest 75\% of pixels in the \textit{detection band} for each OpSim map, and set the LBG magnitude cuts:
\begin{align}
    r\text{-dropouts:} \quad z_\text{cut} &= \min(z_5) \phantom{- 1} \nonumber \\
    i\text{-dropouts:} \quad z_\text{cut} &= \min(z_5) \\
    z\text{-dropouts:} \quad y_\text{cut} &= \min(y_5) \nonumber.
\end{align}
For example, cuts for the baseline~v4.0 survey strategy in year 1 (10) are
\begin{align}
    u\text{-dropouts:} \quad r_\text{cut} &= 23.51 ~ (24.71) \phantom{- 1} \nonumber \\
    g\text{-dropouts:} \quad i_\text{cut} &= 24.49 ~ (25.76) \phantom{- 1} \nonumber \\
    r\text{-dropouts:} \quad z_\text{cut} &= 23.81 ~ (25.13) \phantom{- 1} \\
    i\text{-dropouts:} \quad z_\text{cut} &= 23.81 ~ (25.13) \nonumber \\
    z\text{-dropouts:} \quad y_\text{cut} &= 22.97 ~ (24.19) \nonumber.
\end{align}

Using these values for $m_\text{cut}$, we evaluate Equation~\ref{eq:nz} across the OpSim map (keeping only the deepest 75\% of pixels) and take the mean.
This mean density is the metric used to compare LBG detection for different survey strategies.
Note that while $m_\text{cut}$ is set for the entire sample, $m_5$ varies spatially across the survey footprint due to spatial variation in observing quality.
Even within this deepest 75\%, nonuniformity of LBG number densities due to spatial modulation of $m_5$ will be important for studies of large-scale structure, as this modulation in number density can be mistaken as large-scale structure \citep{weaverdyck2021}.
We will investigate the impact of this modulation on forecasts for LSST cosmology in future work.
We also do not model contamination by low-redshift interlopers, the impact of which are discussed in Section~\ref{sec:conclusion}.

\subsection{Results: LBG Detection}
\label{sec:lbg-results}

\begin{figure*}
    \centering
    \includegraphics[width=0.7\linewidth]{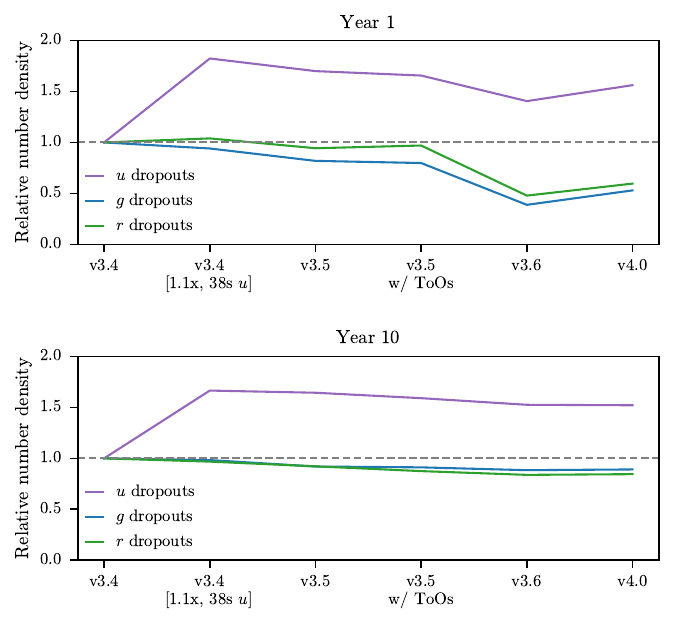}
    \caption{
        Evolution of $u$-, $g$-, and $r$-band dropouts for a series of subsequent survey simulations.
        v3.4 is the baseline simulation after the adoption of the new Ag-Ag-Ag mirror coatings.
        v3.4 [1.1x, 38s $u$] is the same as v3.4, except with 10\% more visits in the $u$ band, with a per-visit exposure time of 38 seconds (compared to 30 seconds in baseline~v3.4).
        v3.5 is the new baseline after the adoption of the [1.1x, 38s $u$] strategy, and includes a few other changes.
        v3.5 w/ ToOs is the same as baseline~v3.5, except with some survey time dedicated to Target of Opportunities.
        Baselines~v3.6 and v4.0 include more realistic simulations of year 1 observatory downtime.
    }
    \label{fig:lbg-timeline}
\end{figure*}

We forecast LBG detection for each survey simulation using the strategy described in Section~\ref{sec:lbg-forecast}, considering first only $u$-band dropouts, which provide LBG samples at $z \sim 3$.
We discuss the higher-redshift $griz$ dropouts at the end of this section.

For the deepest 75\% of the survey footprint in baseline~v3.4, we forecast a $u$-band dropout number densities\footnote{Note these numbers differ from the values plotted in Fig.~\ref{fig:number_density} and printed in Table~\ref{tab:lbg-densities}, as these are for baseline v3.4, while those are for baseline v4.0. This difference is because the $u$-band variants used in this section were produced for baseline v3.4. In Fig.~\ref{fig:number_density} and Table~\ref{tab:lbg-densities}, however, we wish to display number densities corresponding to the most up-to-date survey strategy.} of 69\,deg$^{-2}$ in year 1 and 2113\,deg$^{-2}$ in year 10.
Relative LBG number densities for the range of $u$-band strategy simulations are displayed in Fig.~\ref{fig:lbg-grid}.
Increasing $u$-band depth by increasing the number of $u$-band visits and/or the $u$-band per-visit exposure time results in a greater number density of detected LBGs due to the correspondingly deeper cut that is allowed in the detection band (the $r$ band; cf. Equation~\ref{eq:shallow-cuts}).

Note that for a fixed increase in total $u$-band observing time, increasing the per-visit exposure time has a greater impact than increasing the number of $u$-band visits.
For example, the [1.5x, 30s $u$] and [1.0x, 45s $u$] both correspond to increasing the total $u$-band observing time by 50\%, but the latter strategy with longer per-visit exposures results in a much greater increase in LBG number density.
This is because the Rubin $u$ band is read-noise limited due to the lower sky background at these wavelengths.
Indeed, the impact of a 10\% increase in the number of $u$-band visits is so marginal that the potential for extra accumulated $u$-band depth is not sufficient by the end of year 1 to overcome natural depth variations between different simulation realizations.
Thus, the ``1.1'' column in the left panel of Fig.~\ref{fig:lbg-grid} reports lower LBG number densities than the ``1.0'' column.
This discrepancy also provides an estimate for the precision of our estimated LBG number densities, when considering uncertainties due to natural variations between different simulation realizations.

Another important feature to recognize is that the proposed increases to $u$-band depth have a larger impact in year 1 than in year 10 when comparing relative number of LBGs detected.
Compare, for example, the relative LBG number density increase for the most aggressive $u$-band strategy, which increases the number of $u$-band visits by 20\% and the $u$-band exposure time to 60 seconds.
In year 10, this strategy results in 3.2$\times$ more LBGs detected, a large increase that is, however, much smaller than the corresponding 5.3$\times$ increase in year 1.
This is because year 1 depths are on a steeper part of the LBG luminosity function compared to the deeper year 10 depths (cf. Fig.~\ref{fig:number_density}), so the increase in LBG number density per unit depth is greater in year 1 than in year 10.

\begin{deluxetable*}{lccccc}
    %\tablewidth{\linewidth}
    \tablecaption{Forecast LBG number densities for year 1 (10) of LSST assuming survey strategy baseline~v4.0.}
    \tablehead{
        \colhead{Field} & 
        \colhead{$n_u$} & 
        \colhead{$n_g$} & 
        \colhead{$n_r$} & 
        \colhead{$n_i$} & 
        \colhead{$n_z$}
    } 
    \startdata
        WFD & 110 (3200) & 400 (6300) & 4.4 (230) & 0.19 (9.9) & 0.01 (0.44) \\
        COSMOS & 9400 (13000) & 13000 (27000) & 1100 (2700) & 120 (360) & 0.74 (5.1) \\
        ECDFS & 5400 (13000) & 6900 (23000) & 260 (2300) & 23 (290) & 0.25 (4.3) \\
        ELAIS S1 & 3000 (13000) & 4700 (22000) & 340 (1700) & 29 (210) & 0.17 (2.0) \\
        XMM-LSS & 4400 (11000) & 4400 (18000) & 220 (1600) & 19 (190) & 0.14 (2.3) \\
        EDFS & 2000 (15000) & 5400 (25000) & 280 (2400) & 20 (310) & 0.23 (3.8) \\
    \enddata
    \tablecomments{All number densities are in units of deg$^{-2}$, rounded to 2 significant digits.}
    \label{tab:lbg-densities}
\end{deluxetable*}
%\vspace{-0.9cm}

In October 2024 the Rubin SCOC recommended that LSST adopt the [1.1x, 38s $u$] strategy for the $u$ band \citep{scoc_phase3}.
Note with this strategy and the Ag-Ag-Ag mirror coatings, all bands are deeper than under the previous Al-Ag-Al baseline.
This strategy, marked by the dashed-red boxes in Fig.~\ref{fig:lbg-grid}, was incorporated into the new baseline simulation, v3.5.
Baseline simulations v3.5, v3.6, and v4.0 include other changes, unrelated to $u$-band strategy, that also impact forecast LBG number densities, including dedicating survey time to targets of opportunity\footnote{ToOs are transient events, such as gravitational wave detections \citep{cowperthwaite2019}, that require immediate follow-up observations.} (ToOs), as well as increasing the amount of year 1 observatory downtime to more realistic levels.
The impacts of these changes to survey strategy on the detection of LBGs are shown in Fig.~\ref{fig:lbg-timeline}.

In year 1, the number of detected $u$-band dropouts is seen to increase by a factor of 1.8 under strategy [1.1x, 38s $u$], an increase that is mostly sustained across baseline~v3.5 and the subsequent inclusion of ToOs.
There is a significant decrease in $u$-band dropout detection in baseline~v3.6 due to the more realistic amount of observatory downtime simulated in year 1.
This decrease is somewhat mitigated by the improvements to simulating year 1 downtime implemented in baseline v4.0 (see Section~\ref{sec:opsim}).
A similar pattern is visible for year 10 detections, however the decrease in LBG number densities due to the increased year 1 downtime is far smaller, due to this comprising a much smaller fraction of survey time by year 10.

In both panels we also plot the forecast number density of $g$- and $r$-band dropouts, at redshifts $z \sim 4$ and $z \sim 5$, respectively.
It is seen that the chosen increase to $u$-band survey time results in only very modest reductions in the number densities of these higher-redshift LBGs.
While not shown in the plots, the same is true of $z$- and $y$-band dropouts, at redshifts $z \sim 6$ and $z \sim 7$, respectively.
Number densities for all dropout samples for the baseline~v4.0 survey strategy are shown in Fig.~\ref{fig:number_density}.

Finally, we list projected LBG number densities for simulation baseline v4.0 in Table~\ref{tab:lbg-densities}.
Densities are listed for years 1 and 10, including the WFD survey and all LSST deep fields.
Note that densities are much greater in COSMOS in year 1, compared to the other deep fields, reflecting the early emphasis on building depth in COSMOS for photo-$z$ calibration and low-surface-brightness science.

\section{Conclusions}
\label{sec:conclusion}

This paper used OpSim simulations of the LSST survey, together with simple models of IGM absorption and LBG dropout detection, to evaluate the impact of LSST $u$-band observing strategy on photo-$z$ estimation and detection of LBGs.
We find the following:
\begin{itemize}
    \item Adjusting LSST strategy to increase $u$-band depth has a small, but statistically significant negative impact on photo-$z$ estimation for galaxies at $z < 1.5$. We expect these small changes in performance, however, would be erased by a more realistic treatment of photometry that includes systematic errors, such as aperture corrections.
    \item For galaxies at $z > 1.5$, increasing the $u$-band depth yields a significant improvement in photo-$z$ estimation.
    \item Increasing $u$-band depth has the potential to dramatically increase the number of $z \sim 3$ $u$-band dropouts detected by LSST.
\end{itemize}
These metrics were presented to the Rubin SCOC in June 2024, motivating the SCOC to recommend increasing the number of $u$-band visits by 10\% and the per-visit exposure time to 38 seconds \citep{scoc_phase3}, which has been adopted as the baseline for survey strategy simulations going forward.
With this strategy and the Ag-Ag-Ag mirror coatings, the projected depths in all six bands are deeper than the previous Al-Ag-Al baseline.

With the adoption of this $u$-band strategy, our metrics indicate the following:
\begin{itemize}
    \item The outlier fraction, standard deviation, and bias of $z < 1.5$ galaxies will \textit{increase} by 20\%, 5\%, and 50\%, respectively.
    These values are, however, small in absolute magnitude, and all three metrics remain well below the LSST requirements.
    \item The outlier fraction, standard deviation, and bias of $z > 1.5$ galaxies will \textit{decrease} by up to 28\%, 10\%, and 25\%, respectively.
    As all three metrics are significantly larger at high redshifts, these also represent large improvements in absolute value.
    \item The number density of $u$-band dropouts will increase by 82\% in year 1 and 67\% in year 10. Forecast number densities for $ugriz$-dropouts in the LSST WFD and DDFs are listed in Table~\ref{tab:lbg-densities}.
\end{itemize}
This represents significant gains for science with high-redshift galaxies.
For example, the $u$-band dropout sample forecast to have number densities of 110~deg$^{-2}$ (3200~deg$^{-2}$) in year 1 (10) will enable measurement of the cross correlation with Simons Observatory CMB lensing at an SNR of 90 (160) \citep{wilson2019,ade2019}.

Given the high SNR forecast for our observables, the precision of cosmological constraints will ultimately be limited by our control of systematic errors.
One of the most important sources for systematic error to consider is photo-$z$ contamination from low-redshift interlopers. 
Galaxies with strong Balmer/4000\,\AA\, breaks, dusty galaxies, emission line galaxies (ELGs), and low-temperature dwarf stars all contaminate LBG dropout samples because they have colors that mimic the Lyman break \citep{stanway2008, reddy2008, vulcani2017, ono2018}.
Indeed, \citet{ruhlmann-kleider2024} and \citet{payerne2024} found the purity of $u$-band dropout samples depends strongly on $u$-band depth, with the number of low-redshift interlopers dropping by 38\% as $u$-band depth increases from 24.5 to 25.5.
Furthermore, deep $u$-band imaging is valuable for suppressing interlopers in high-redshift dropout populations by rejecting sources with non-negligible flux in wavelengths bluer than the supposed dropout band \citep{vulcani2017}.
Thus, while this paper considers only detected number densities of true LBGs, greater $u$-band depth will also increase purity for LBG dropout samples at all redshifts.
The deep photometry and data at other wavelengths available in the LSST DDFs will be invaluable for characterizing interloper populations present in LSST LBG dropout samples.
Careful study and calibration of these populations will be necessary to enable precision cosmology with LBGs detected by LSST \citep{petri2025}.

Finally, we note that increasing the $u$-band depth of LSST imaging enhances synergies with the proposed DESI-II survey \citep{Schlegel2022desi2}, the extension of Dark Energy Spectroscopic Instrument (DESI) \citep{DESI2016whitepaper} which aims to map the 3D matter distribution in the $2 < z < 4$ universe using spectroscopic samples of LBGs and Lyman-alpha emitters (LAEs), enabling tests of cosmological models in the matter-dominated era.
DESI-II plans to use LSST year 2 catalogs to select candidate LBGs and LAEs for spectroscopic follow-up.
As previously discussed, increasing LSST $u$-band depth will increase the number density and purity of these candidate samples, increasing the efficiency of the DESI-II survey.

The code to produce the plots in this paper is available on GitHub\footnote{\url{https://github.com/jfcrenshaw/u-band-strat}} and Zenodo \citep{u-band-strat-notebooks}.
The code for the LBG models described in Section \ref{sec:lbg-pop-model} is published as the Python package \texttt{lbg\_tools}\footnote{\url{https://github.com/jfcrenshaw/lbg_tools}}, which is available on PyPI\footnote{\url{https://pypi.org/project/lbg-tools/}};
this paper was produced using version v1.4.1 \citep{lbg_tools}.
The Rubin OpSim simulations are stored on Zenodo \citep{opsim3p4,opsim3p5,opsim3p6,opsim4p0}.

\begin{acknowledgements}

We thank Yoshiaki Ono for providing HSC completeness curves.
This paper has undergone internal review by the LSST Dark Energy Science Collaboration.
The authors thank the internal reviewers, Sam Schmidt and Rebecca Chen, for their valuable comments.
The authors also thank the anonymous journal referee for their excellent and comprehensive review.

JFC acknowledges support from the U.S. Department of Energy, Office of Science, Office of High Energy Physics Cosmic Frontier Research program under award number DE-SC0011665.  
EG acknowledges support from the U.S. Department of Energy, Office of Science, Office of High Energy Physics Cosmic Frontier Research program under award number DE-SC0010008 and from an IBM Einstein fellowship for his sabbatical at IAS during the completion of this manuscript. 
AIM acknowledges the support of Schmidt Sciences.
BL is supported by the Royal Society through a University Research Fellowship.

The DESC acknowledges ongoing support from the Institut National de Physique Nucl\'eaire et de Physique des Particules in France; the Science \& Technology Facilities Council in the United Kingdom; and the Department of Energy and the LSST Discovery Alliance in the United States.
DESC uses resources of the IN2P3 Computing Center (CC-IN2P3--Lyon/Villeurbanne - France) funded by the Centre National de la Recherche Scientifique; the National Energy Research Scientific Computing Center, a DOE Office of Science User Facility supported by the Office of Science of the U.S.\ Department of Energy under contract No.\ DE-AC02-05CH11231; STFC DiRAC HPC Facilities, funded by UK BEIS National E-infrastructure capital grants; and the UK particle physics grid, supported by the GridPP Collaboration.
This work was performed in part under DOE contract DE-AC02-76SF00515.

This research has made use of NASA's Astrophysics Data System and The Software Citation Station\footnote{\url{https://www.tomwagg.com/software-citation-station/}} \citep{software-citation-station-paper, software-citation-station-zenodo}.

\end{acknowledgements}

\begin{contribution}

JFC designed the LBG metrics, performed the analysis, and wrote the majority of the paper.
BL proposed the idea to create LBG survey strategy metrics, developed prototypes, and provided feedback on the text.
MG produced the photo-$z$ metrics and provided feedback on the text.
CP provided significant feedback on all aspects of the text and contributed connections with DESI.
AJC suggested approaches and designed the CMNN photo-$z$ estimator.
EG suggested approaches and provided feedback on the text. 
TK provided input on aspects of LBG cosmology and connections to DESI.
AIM contributed to the photo-$z$ metrics.
JN suggested approaches and provided feedback on metrics.
MR provided feedback on LBG cosmology, including relevance to cluster cosmology.

\end{contribution}

\software{
    astropy \citep{astropy:2013, astropy:2018, astropy:2022},
    CMNN \citep{graham2018, graham2020},
    HEALPix \citep{gorski2005} and healpy \citep{zonca2019},
    jupyter \citep{jupyter},
    MAF \citep{maf},
    matplotlib \citep{matplotlib},
    numpy \citep{numpy},
    pandas \citep{pandas,pandas-software},
    Python \citep{python},
    scipy \citep{scipy},
}

\appendix

\section{Comments on $\beta_\text{UV}$ dependence\\of IGM increments $\Delta \MakeLowercase{m}_\text{\MakeLowercase{w}IGM}$}
\label{sec:beta-dependence}

\begin{figure*}
    \centering
    \includegraphics[width=0.89 \linewidth]{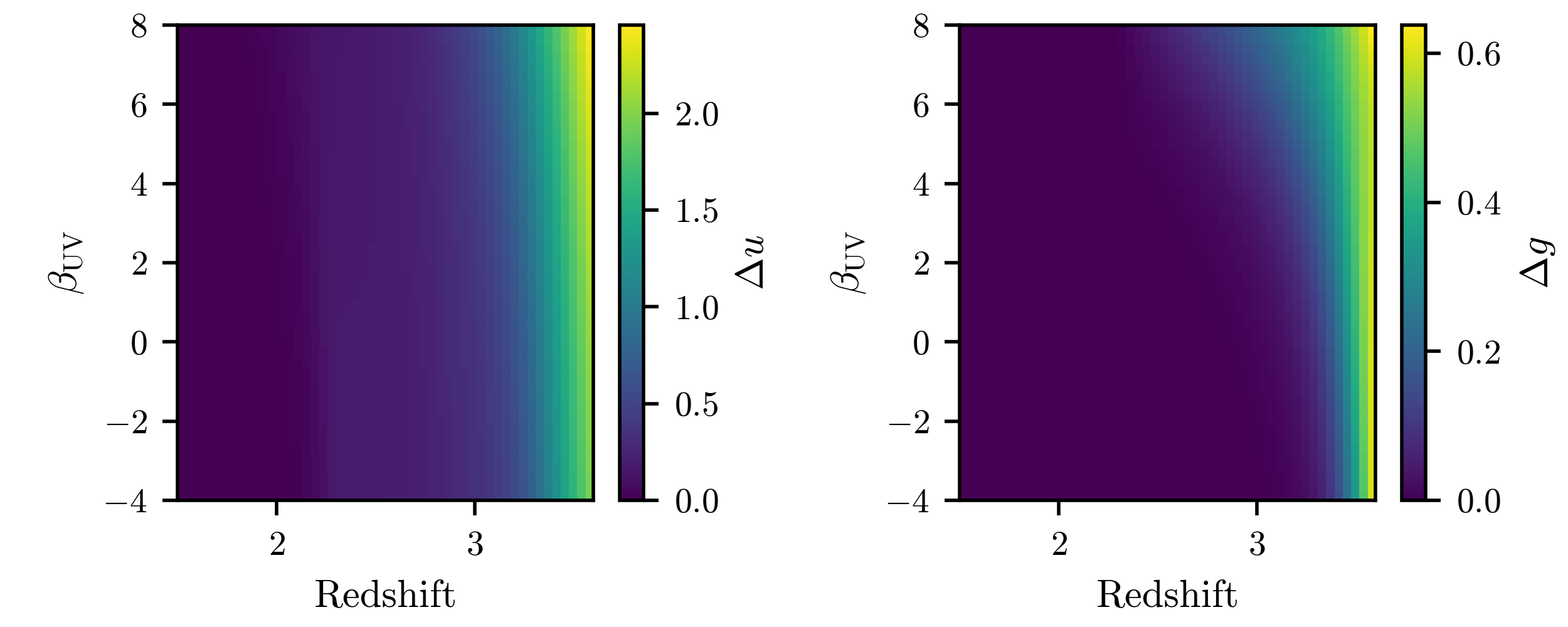}
    \caption{
        IGM magnitude increments for the Rubin $u$ and $g$ bands, as a function of source redshift and UV slope.
    }
    \label{fig:increment_maps}
\end{figure*}

Note the shapes of IMG transmission curves in Fig.~\ref{fig:igm} provide some insight into the effect of the UV slope.
When the Lyman-alpha forest starts to redshift into a band, bluer spectra will have a larger IGM correction, $\Delta m_\mathrm{IGM}$, as they have greater flux in the wavelength range impacted by extinction.
However, as IGM extinction redshifts farther into the band, redder spectra will have a larger correction, as the IGM has a greater optical depth at high redshift, corresponding to longer wavelengths.
You can see this from the deeper troughs on the right side of the IGM extinction curves in Fig.~\ref{fig:igm}.
However, once the Lyman limit redshifts into the band, bluer spectra will once again have larger corrections as the IGM is far more opaque at wavelengths below the Lyman limit.

Fig.~\ref{fig:increment_maps} plots $\Delta m_\mathrm{IGM}$ for the $u$ and $g$ bands as a function of both redshift and UV slope.
It is difficult to see the structure described above, however, indicating that these effects are relatively small.

\section{Fitting the $\beta_\text{UV}$ model}
\label{sec:uv-slope-fit}

\begin{figure}
    \centering
    \includegraphics[width=0.88\linewidth]{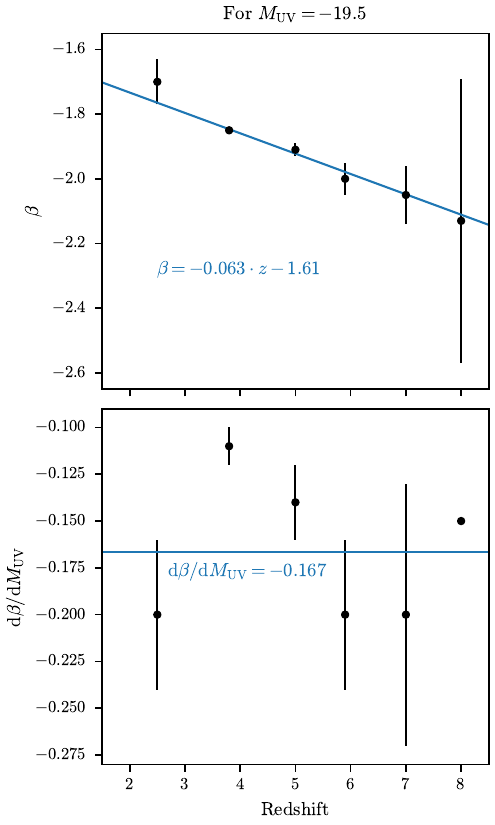}
    \caption{
         Model for evolution of $\beta_\text{UV}$, fit to data from Table~3 of \citep{bouwens2014}.
    }
    \label{fig:uv_slope_fit}
\end{figure}

To model the UV spectra of LBGs we use Hubble Space Telescope data from  \citet{bouwens2014}.
Specifically, we use the linear fit parameters listed in Table~3 for six tomographic redshift bins between $2.5 \lesssim z \lesssim 8.0$.
These parameters, as a function of mean redshift, are plotted in Fig.~\ref{fig:uv_slope_fit}.

The parameter $\beta_\text{UV}$ at $M_\text{UV} = -19.5$ shows a clear linear trend with redshift, so we fit a linear model, yielding the relation
\begin{align}
    \beta_\text{UV} |_{M_\text{UV} = -19.5} = -0.063 z - 1.61.
\end{align}
The parameter $\mathrm{d}\beta_\text{UV}/\mathrm{d}M_\text{UV}$ does not show a clear trend, so we simply take the average:
\begin{align}
    \left\langle \frac{\mathrm{d}\beta_\text{UV}}{\mathrm{d}M_\text{UV}} \right\rangle = -0.167.
\end{align}
Together, these two relations yield the bilinear model
\begin{align}
    \beta_\text{UV}(M, z) = -0.167(M + 19.5) - 0.063 \, z - 1.61,
\end{align}
which is also printed as Eq.~\ref{eq:beta-model}.

\section{Details of the LBG\\Completeness Model}
\label{sec:completeness_model}

The LBG completeness models of \citet{malkan2017,ono2018,harikane2022} were calibrated using synthetic source injection (SSI). 
That is, synthetic galaxy images with a variety of spectral types, redshifts, and intrinsic magnitudes were injected into real images, and these images were processed using the usual science pipelines used for these studies.
The fraction of true LBGs injected into the images were then compared with the number of LBGs that were detected in the images and then passed the corresponding color cuts.
The completeness was then calculated in bins of redshift and apparent magnitude.

Estimates for $u$-dropout completeness come from \citet{malkan2017}.
This study used photometry from the Subaru Deep Field, and detected LBGs using $z$-band imaging with 5$\sigma$ depth\footnote{
This PSF depth is estimated from the 3$\sigma$ depth in a 2'' aperture listed in Table~1 of \citet{malkan2017}, using the formula
\begin{align*}
    z_5 &= z_3 + 2.5 \log_{10}\frac{3}{5}  - 2.5 \log_{10}\sqrt{\frac{A_\text{PSF}}{A_\text{Aperture}}},
\end{align*}
where the latter term is the ratio of the PSF and aperture areas.
} approximately 26.8.
While we use these same completeness curves, we assume $u$-dropout detection occurs in the $r$ band, justified by the relatively flat spectra of LBGs above the Lyman-break.
The LSST $r$ band has a 5$\sigma$ depth $\sim 25.7$ in year 1 and $\sim 26.8$ in year 10, which is well matched to the $z$-band depth from the Subaru Deep Field.
Estimates for $griz$-dropout completeness come from \citet{ono2018,harikane2022}, which use photometry from the HSC SSP Wide, Deep, and Ultradeep fields.
These fields span a range of 5$\sigma$ depths that bracket the expected LSST year 1 and 10 depths.

In every case, we use the strategy of \citet{harikane2022} to ``rescale'' completeness estimates to imaging of different depths.
That is, we take the input grid of redshift, apparent magnitude, and completeness values, $\{z, ~ m, ~ C\}$, and subtract the 5$\sigma$ depth from the apparent magnitudes:
\begin{align}
    \{z, ~ m, ~ C\} \to \{z, ~ m - m_5, ~ C\}.
\end{align}
Completeness, therefore, is modeled as a function of redshift and \emph{magnitude relative to the 5$\sigma$ depth in the detection band}.
We then linearly interpolate (and extrapolate) using this grid, with completeness values clipped to the range $[0, 1]$ to keep completeness for very bright (faint) galaxies $\leq 1$ ($\geq$ 0). 
We modify the input completeness curves so they are unimodal with respect to redshift, and so that interpolated/extrapolated completeness values decrease monotonically with magnitude.
This improves the behavior of extrapolation beyond the input calibration grid.
A set of interpolated completeness curves is displayed in Fig~\ref{fig:completeness}.

Note we do not expect these completeness models to be correct \emph{in detail} for LSST.
For example, the $u$-band depth in \citep{malkan2017} is closer in depth to the detection bands than the LSST $u$ band is to the $r$ band.
The $u$-dropout completeness, therefore, may be lower for LSST.
We note, however, the $u$-band completeness curves in Fig.~\ref{fig:completeness} are qualitatively similar to the completeness curves for the higher-redshift dropouts, suggesting this model still reasonably captures the scaling of completeness with LBG magnitude relative to survey depth.
Furthermore, we note that our forecast for $u$-dropout number densities is of a similar order of magnitude as, but a factor of few smaller than, other forecasts for detection of $u$-band dropout detection in LSST \citep{wilson2019,petri2025}.

In summary, we expect our LBG population model provides reasonable estimates for the order of magnitude of LBGs that LSST will detect, reasonable redshift ranges for these detections, and, \emph{most importantly}, a reasonable scaling with LBG magnitude relative to the 5$\sigma$ depth of the imaging.
Quantifying completeness in detail for LSST will be vital for cosmology with LBGs detected by Rubin, however we do not think these details impact the conclusions of this paper.

Finally, we note that Equation~\ref{eq:nz} uses absolute magnitudes at 1500\,\AA, $M$, rather than apparent magnitudes, $m$.
Equations~\ref{eq:lbg_flux_intr}-\ref{eq:apparent-mag}, however, provide a fully-determined analytic model for $m = m(M, z)$.
These equations can be numerically inverted to provide a fully determined model for $M = M(m, z)$.

\bibliography{references, references_supplement}{}
\bibliographystyle{aasjournal}

\end{document}